\documentstyle[epsfig,12pt]{article}
\begin{document}
{\Large \bf Charmonium dissociation in hot and dense matter}
\footnote{Talk at the Conference ``Progress in Heavy Quark Physics'', 
Rostock, Germany, Sept. 20-22, 1997}
\\[2mm]
{\it K.~Martins $^{a,b}$ and D.~Blaschke $^a$ }\\[2mm]
{\small $^a$University of Rostock, $^b$ Bio Art Products GmbH Rostock}\\[2mm]
{\normalsize\bf Abstract}\\
We investigate $J/\psi$ suppression in ultrarelativistic heavy ion collisions
within a generalized {\sc Glauber} model and compare results of a hadronic
comover scenario with those of a quark plasma 
scenario where conditions for {\sc Mott} dissociation of charmonium are 
fulfilled locally at the short time scales before hadronization.
We study possible consequences for the theoretical description to be drawn 
from the observation of ``anomalous'' thresholds in the $J/\psi$ suppression 
pattern by the NA50 collaboration. 
%and make suggestions for future experiments.

\section{Generalized {\sc Glauber} model for $AB$ collisions} 

Recently, the NA50 collaboration at CERN has observed an ``anomalous'' 
threshold effect for $J/\psi$ production in $Pb-Pb$ collisions at
158 $A$ GeV/c [1,2].  After the debate whether the $J/\psi$ suppression itself
already signals quark-gluon plasma formation [3] or may be explained
by hadronic absorption [4], the present observation seems to support a
plasma scenario predicting anomalous $J/\psi$ suppression thresholds
to occur due to the {\sc Mott} effect [5] which excludes charmonium
bound state formation in a medium at densities and/or temperatures
just above those which have been reached in previous $AB$ collisions
at 200 $A$ GeV/c 
($A=^{16}\hspace{-2pt}O,~^{32}\hspace{-2pt}S,~B=^{238}\hspace{-2pt}U$).  
In order to explain
both the normal and the anomalous $J/\psi$ suppression pattern in $AB$
collisions quantitatively, generalizations of the {\sc Glauber} model have
been considered which include density thresholds for charmonium
production [6,7]. On the other hand, it has been checked whether more
conventional scenarios of $J/\psi$ absorption by comoving matter can
give a reasonable description of the $Pb-Pb$ data [8,9].  In this
contribution we report on the present status of the interpretation of
the new data for the $E_T$ dependence of $J/\psi$ suppression within a
generalized {\sc {\sc Glauber}} model approach which besides of charmonium
attenuation by projectile and target nucleons also includes absorption
by comoving hadronic and/or partonic matter as produced in the course of the
collision.  Within this approach, the survival probability of a
charmonium state produced in a collision of the nuclei $A$ and $B$ at
impact parameter $b$ is given by [10]
\begin{eqnarray}
S(b)&=& \left[{d\sigma_{prod} \over d\, b }\right]^{-1} \int_0^\infty
        d^2{\bf b}_A {d^2\sigma _{prod} \over d\,b d\,b_A}
        P^{abs}_{nucl,A}({\bf b}_A)P^{abs}_{nucl,B}({\bf b}-{\bf b}_A)
        P^{abs}_{com}({\bf b, b}_A)~,
\end{eqnarray}
where the absorption by projectile and target nucleons with density
distributions $\rho_i(z_i,{\bf b})$ is given by
\begin{eqnarray}
P^{abs}_{nucl,i}({\bf b})&=& \int_{-\infty}^\infty d\, z_i
\rho_i(z_i,{\bf b})\left[ 1- \int_{z_i}^\infty d\, z \rho_i(z,{\bf b})
\sigma_{abs}^{(c\bar c) N}\right]^{i-1}~;~~i=A, B~,
\end{eqnarray}
and $\sigma_{abs}^{(c\bar c) N}$ is a phenomenological cross section
adjusted in order to describe the mass number dependence of charmonium
hadroproduction in $pA$ and $\pi A$ collisions where no dense medium
is expected to be formed [4,9].  The impact parameter dependence can be
translated to an $E_T$ dependence using the $E_T - b$ correlation
function which is also obtained within the {\sc Glauber} model approach from
a fit to the experimental dimuon $E_T$ distribution.  The additional
absorption by hot, dense matter produced in the central region of an
ultrarelativistic nucleus-nucleus collision is accounted for by
\begin{eqnarray}
\label{pcom}
P^{abs}_{com,h}({\bf b, b}_A)&=& \exp\left[-\int_{t_0}^{t_f} d~t~
\tau^{-1}({\bf b, b}_A,t)\right]~,
\end{eqnarray}
where $\tau({\bf b, b}_A,t)$ is the relaxation time for the initial
distribution of $c\bar c$ pairs due to inelastic collisions with
hadrons or partons of the hot and dense medium.  We present results of
model calculations with and without a critical threshold behaviour in
the relaxation time in order to discuss whether or not the
experimental signal of anomalous $J/\psi$ suppression can be
interpreted as a signature of quark-gluon plasma formation.

\section{Charmonium dissociation in hadronic matter}

In dense hadronic matter charmonium formation is suppressed by charm 
dissociation into open charm hadrons.  
We have calculated the cross sections $\sigma_h$
of such quark rearrangement (string-flip) processes using a confining
quark potential model [11] and obtained the relaxation time
$~\tau_h=\langle\sigma_h v_{rel} n_h\rangle^{-1}~$ 
including a parametric dependence on the hadron properties 
(masses and r.m.s. radii). 
In Fig.~1 we contrast a conventional description in terms of nuclear absorption
(full lines, $\sigma_{abs}=7.3$ mb)  with different hadronic comover absorption
scenarios: dotted lines-
$\sigma_h={\rm const}$, $n_h(t)=n$, dashed lines-   
$\sigma_h={\rm const}$, $n_h(t)=n~t_0/t$ and dash-dotted lines-
$\sigma_h={\rm const}~n$, $n_h(t)=n~t_0/t$. 
Our results [12] for the additional absorption due to comoving hadrons 
show a continuous dependence on the impact parameter. 
No acceptable description of both sets of data 
($S-U$ and $Pb-Pb$) is possible and a hadronic comover scenario has to be 
disfavoured, see also [9].
\begin{figure}
\begin{minipage}[h]{100mm}
{\psfig{figure=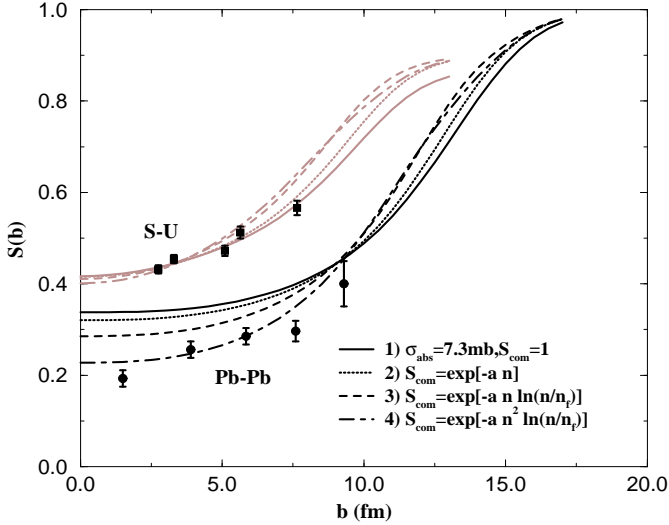,height=8cm,width=10cm,angle=0}}
\end{minipage}\hfill
\begin{minipage}[h]{50mm}
%~\vfill
\vspace*{-5mm}
\caption{ $J/\psi$ survival probability vs. impact parameter for S-U and
for Pb-Pb collisions [1]. No simultaneous description of both data sets is 
possible within the hadronic comover scenario even with a strong density 
dependence of the charmonium dissociation cross section (curve 4)), see text.}
\vfill ~
\end{minipage}
\vspace*{-1cm}
\end{figure}
%\\[2mm]
%\parbox{10cm}
%{\psfig{figure=martins/Nukl-Com-Vgl.eps,height=8cm,width=10cm,angle=0}}
%\hfill
%\parbox{6cm}
%{\small Fig.\ 1. $J/\psi$ survival probability vs. impact parameter for S-U and
%for Pb-Pb collisions [1]. No simultaneous description of both data sets is 
%possible within the hadronic comover scenario even with a strong density 
%dependence of the charmonium dissociation cross section (curve 4)).}

\section{Mott dissociation of $c\bar c$ in partonic matter}

According to the {\sc Mott} effect, above critical densities characteristic 
for the bound states under consideration, these states vanish from the spectrum
since they merge the continuum of scattering states. 
In the case of charmonium, quark 
rearrangement processes occur which result in a dissociation of the heavy 
flavour [13] when critical parton densities $n_{\sc Mott}^{c\bar c}$ are 
exceeded.
For a $c\bar c$ pair born into a dense partonic environment, a {\sc Mott} 
criterion has been formulated in [7] which accounts for the proper time 
dependence of both the size of the $c\bar c$ wave function and the parton 
density [14] at given impact parameter and {\sc Lorentz} factor $\gamma$ in 
the parton center-of-mass system, see Fig. 2.
\begin{figure}
\begin{minipage}[t]{100mm}
{\psfig{figure=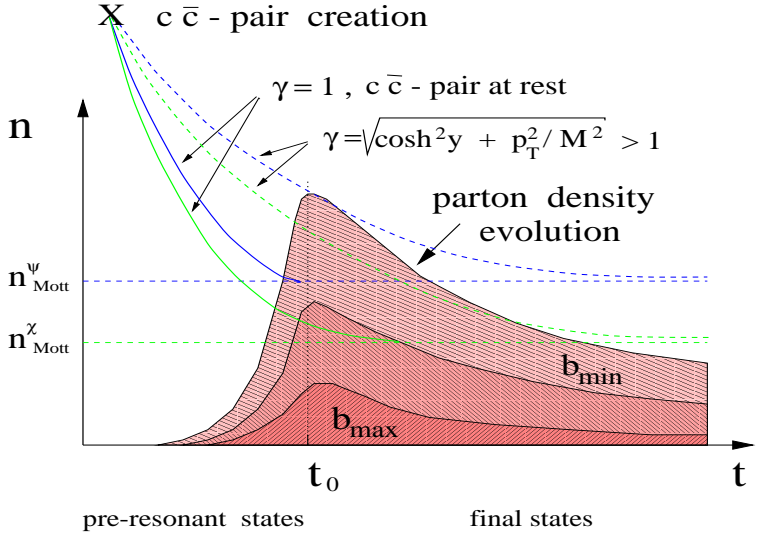,height=10cm,width=7cm,angle=-90}}
\end{minipage}\hfill
\begin{minipage}[t]{50mm}
~\vfill
\caption{Proper-time dependence of parton density (hatched regions) and 
critical Mott-densities (pre-resonant states: lines joining the $c \bar c$ 
creation vertex, final states: horizontal lines).}
\vfill~
\end{minipage}
\vspace*{-5mm}
\end{figure}
Accounting for a 30\% feeding of the $J/\psi$ channel from $\chi$ decays, we 
obtain for the additional absorption due to {\sc Mott} dissociation of 
charmonium in dense partonic matter 
\begin{eqnarray}
P^{abs}_{com,{\sc Mott}}({\bf b, b}_A)&=&
0.3~\Theta[\gamma^2 n_{\sc Mott}^\chi-n({\bf b, b}_A)] +
0.7~\Theta[\gamma^2 n_{\sc Mott}^\psi-n({\bf b, b}_A)]~,
\end{eqnarray}
$n({\bf b, b}_A)=AT_A({\bf b}_A)(1-(1-\sigma_{inel}T_B({\bf b-b}_A))^B)
+BT_B({\bf b-b}_A)(1-(1-\sigma_{inel}T_A({\bf b}))^A)$
being the density of participants in the transverse plane  
and $T_{A,B}({\bf b})$ the normalized nuclear thickness functions. 
\begin{figure}[h]
\vspace{-10mm}
\psfig{figure=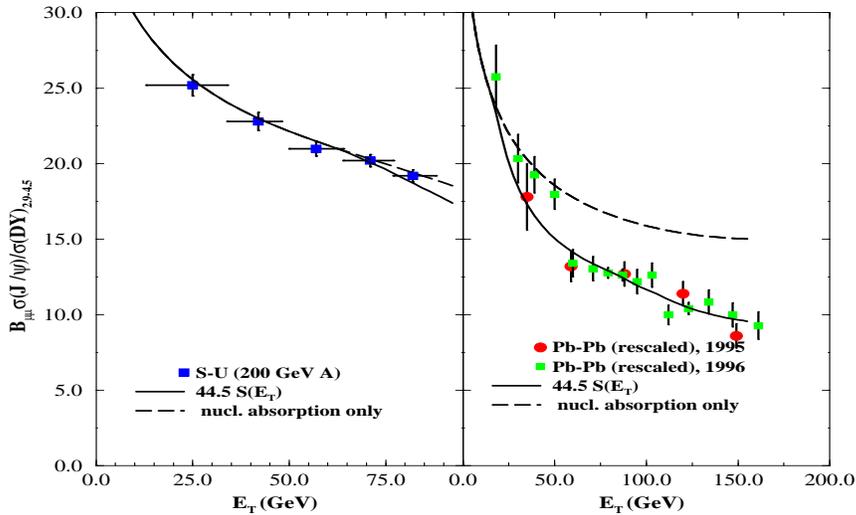,height=13cm,width=8cm,angle=270}
\caption{ Suppression ratio for $J/\psi$ in S-U (left panel) and 
Pb-Pb (right panel) collisions.}
\end{figure}
From the r.m.s. transverse radii of $\chi$ and $J/\psi$ we expect that 
$ n_{\sc Mott}^\psi \approx 2~n_{\sc Mott}^\chi$ and the parameter 
$n_{\sc Mott}^\chi$ is chosen such that the threshold for anomalous suppression
which is interpreted as a consequence of the {\sc Mott} effect is reproduced. 
Our results are shown in Fig. 3 and demonstrate that a threshold behaviour can
indeed be obtained for the $Pb-Pb$ collisions whereas it is not present in the
$S-U$ case. However, the most recent data of the 1996 run [2] exhibit
a much steeper decrease of the $J/\psi$ suppression ratio than can be explained
within a generalization of the {\sc Glauber} model even by the extreme 
{\sc Mott} scenario as presented here and in [7]. 
\section{Conclusions}

We have analyzed data for the $E_T$ dependence of $J/\psi$ suppression in $S-U$
and $Pb-Pb$ collisions at CERN-SpS within a generalized {\sc Glauber} model 
approach which accounts for the fact that after its creation the $c\bar c$ pair
suffers not only normal absorption by projectile and target nucleons but also
additional absorption while traversing the dense hadronic and/or partonic 
matter which is formed in the collision.
It has been shown that a hadronic comover scenario cannot explain both the 
$S-U$ and the $Pb-Pb$ data since it has a continuous centrality dependence.
The scenario of {\sc Mott} dissociation in dense partonic matter when 
implemented in the {\sc Glauber} approach exhibits a threshold behaviour with 
increasing centrality of the collision and seems to describe both the $S-U$ 
and the 1995 $Pb-Pb$ data. The new 1996 $Pb-Pb$ data, however, show a sudden 
drop in the $J/\psi$ production above $E_T\sim 50$ GeV which, when established 
in the final analysis of the data, indicates the transition to a new regime in 
the  $Pb-Pb$ collisions that can no longer be modeled adequately within a 
{\sc Glauber} type approach.
  
Studies which implement the change of time scales due to a quark-gluon-plasma 
phase transition [5,15] on the other hand seem to have the potential to 
explain the observed anomalous threshold behaviour of $J/\psi$ suppression. 
These theoretical studies have to be further developed by improving the 
microscopic description of charmonium dissociation kinetics and by combining 
them with numerical simulations of $AB$ collisions.  
Further experimental studies with higher statistics and also with inverse 
kinematics should be performed in order to settle the details of the anomalous 
suppression pattern as a possible signal of quark-gluon-plasma formation in 
heavy-ion collisions.  
\\[5mm]
\rule{\textwidth}{1pt}

\noindent
%{\small$^\dagger$ This is an odd footnote.}\\
\normalsize
1.\ M.~Gonin {\em et al.} (NA50 Collaboration),  
%       in {\it Proceedings Quark Matter '96}, Eds. P. Braun-Munzinger, 
%        H.-J. Specht, R. Stock and H. St\"ocker, 
        Nucl. Phys. {\bf A 610} (1996) 404c.\\
2.\ E. Scomparin (NA50 Collaboration), talk at this conference.\\ 
3.\ T.~Matsui and H.~Satz, Phys. Lett. {\bf 178 B} (1986) 416.\\
4.\ C.~Gershel and J.~H\"ufner, Z. Phys. {C 56} (1992) 171.\\
5.\ D.~Blaschke, Nucl. Phys. {\bf A 525} (1991) 269c.\\
6.\ J.-P.~Blaizot and J.-Y.~Ollitrault, Phys. Rev. Lett. {\bf 77}
    (1996) 1703. \\
7.\ K. Martins and D. Blaschke, in {\it Proceedings Hirschegg '97 
     ``QCD Phase Transitions''}, p. 293; nucl-th/9702044.\\
8.\ S.~Gavin and R.~Vogt, Nucl. Phys. {\bf A 610} (1996) 442c.\\
9.\ D.~Kharzeev, C.~Louren\c{c}o, M.~Nardi and H.~Satz, Z. Phys. {\bf C 74}
    (1997) 307.\\
10.\ C.~Y.~Wong, Phys. Rev. Lett. {\bf 76} (1996) 196.\\
11.\ K.~Martins, D.~Blaschke and E.~Quack, Phys. Rev. {\bf C 51} 
        (1995) 2723.\\
12.\ K.~Martins, Ph.D. Thesis, Rostock (1997), unpublished.\\
13.\ G.~R\"opke, D.~Blaschke and H.~Schulz, Phys. Rev. {\bf D 38} 
     (1988) 3589.\\
14.\ K.~Geiger and B.~M\"uller, nucl-th/9707048.\\
15.\ D.~Kharzeev, M.~Nardi and H.~Satz, hep-ph/9707308.

\end{document}